\documentclass[11pt]{article}

\usepackage[preprint]{latex/acl}

\usepackage{times}
\usepackage{latexsym}
\usepackage[T1]{fontenc}
\usepackage[utf8]{inputenc}
\usepackage{microtype}
\usepackage{inconsolata}
\usepackage{graphicx}
\usepackage{listings}
\usepackage{url}
\usepackage{enumitem}
\usepackage{booktabs}
\usepackage{pifont}
\usepackage{placeins}
\usepackage{float}
\newcommand{\cmark}{\ding{51}}
\newcommand{\xmark}{\ding{55}}

\title{Edhivo: An AI-Native Platform for Scalable Assessment and Instant Feedback in Higher Education}

\author{Moraish Kapoor \\
  Stony Brook University \\
  \texttt{moraish.kapoor@stonybrook.edu} \\\And
  Jiawei Zhou \\
  Stony Brook University \\
  \texttt{jiawei.zhou.1@stonybrook.edu} \\}

\begin{document}
\maketitle
\begin{abstract}
Scaling assessment in higher education faces two intertwined bottlenecks: the cognitive burden of \emph{authoring} assessments and the inability to \emph{semantically evaluate} open-ended responses, which together create a persistent feedback gap that hinders student learning outcomes. To address these challenges, we present Edhivo, an AI-native pedagogical platform that addresses both bottlenecks through a vertically integrated system for content delivery, assessment, and reporting. Edhivo employs a Dual-LLM pipeline: one model generates context-aware assessments for both multiple-choice and open-ended questions, from instructor-provided materials, while a second model performs multi-criteria, rubric-based evaluation of student responses, enabling near-instant qualitative feedback at multiple hierarchical levels. All AI interactions are source-grounded to reduce hallucination risk, and a Human-in-the-Loop review process preserves instructor authority. We demonstrate Edhivo's efficacy through a pilot deployment at Stony Brook University in a real classroom environment in 2026 spring, where it supports high-frequency assessment creation and evaluation for a cohort of 36 students throughout a full semester, significantly reducing authoring effort and feedback latency while enabling novel pedagogical designs centered on active learning.\footnote{Edhivo platform is available at \url{https://sbu.edhivo.com/} and is ready to use for students and instructors. A system demonstration video is available at \url{https://www.youtube.com/watch?v=rOvfwFM-eW4}.}
\end{abstract}

\section{Introduction}
\label{sec:intro}

With the rapid expansion of enrollment in higher education, instructors face two intertwined bottlenecks that hinder effective learning outcomes. First, the \emph{authoring burden} of creating high-quality assessments---reviewing materials, formulating questions, and drafting answer keys---is a cognitively demanding and time-intensive process. Second, existing systems lack the \emph{semantic depth} to evaluate open-ended student responses, restricting scalable assessment to shallow question types such as multiple-choice and creating a persistent \emph{feedback gap} that prevents students from correcting misconceptions while the material is fresh \cite{hattie2007power, shute2008focus}. Traditional Learning Management Systems (LMS), such as Canvas \cite{canvas2024} or Blackboard \cite{blackboard2024}, primarily function as administrative hubs and treat AI as a secondary plugin rather than a core pedagogical component, leaving both bottlenecks largely unaddressed \cite{black1998inside}. Meanwhile, emerging AI-powered study aids such as NotebookLM \cite{notebooklm2024} and Turbo.ai \cite{turboai2024} offer source-grounded content generation for self-study but lack instructor-led workflows, cohort reporting, and the human-in-the-loop controls required for formal classroom environments where instructors must be able to review and modify generated content.

This paper introduces Edhivo, which tackles both bottlenecks through a vertically integrated, AI-native architecture. Rather than bolting AI onto an existing LMS, Edhivo places a Dual-LLM Pipeline at its core: a generation model produces context-aware MCQ and open-ended assessments from instructor-provided materials, while a separate evaluation model scores responses across correctness, completeness, and relevance using instructor-defined ground truth. A Human-in-the-Loop review process preserves instructor authority, and all AI interactions remain source-grounded---constrained to uploaded course materials---to reduce hallucination risk.

We demonstrate the system's efficacy through a pilot deployment at Stony Brook University within a real classroom environment,\footnote{See \url{https://joezhouai.com/llm-course-26/} for the course information.} where a novel pedagogy of student-led assessment creation and evaluation is designed as a center piece of the learning process. Our results indicate that Edhivo significantly eases assessment authoring, reduces feedback latency, and lowers instructor workload.

\section{Related Work}
The landscape of educational technology can broadly be characterized by three categories: legacy administrative systems, synchronous engagement tools, and emerging AI-powered study aids \cite{kasneci2023chatgpt}. While these categories increasingly overlap, each addresses a different slice of the instructional workflow. Edhivo integrates and extends these functionalities into a unified, AI-native classroom solution specialized for assessment-based learning.
Table~\ref{tab:comparison} summarizes the key capability differences, with details below.

\paragraph{Traditional LMS} Canvas \cite{canvas2024} and Brightspace \cite{brightspace2024} excel at content hosting and gradebook management but treat AI as a plugin; their assessment models remain largely summative and manual, leaving the feedback gap unaddressed for open-ended responses.

\paragraph{Interactive Response Systems} Kahoot \cite{kahoot2024} and Slido \cite{slido2024} enable gamified, real-time engagement but are limited to atomic question types (primarily MCQ) and lack persistent course-content grounding for qualitative feedback.

\paragraph{AI-Native Study Aids} NotebookLM \cite{notebooklm2024} and Turbo.ai \cite{turboai2024} pioneered source-grounded AI for self-study but lack instructor-led workflows, cohort reporting, and rubric-based evaluation required for formal education.

\begin{table}[t]
\centering
\small
\renewcommand{\arraystretch}{1.2}
\resizebox{\columnwidth}{!}{
\begin{tabular}{lcccc}
\toprule
\textbf{Capability} & \textbf{Canvas} & \textbf{Kahoot} & \textbf{NbLM} & \textbf{Edhivo} \\
\midrule
AI-Native Eval. & \xmark & \xmark & \xmark & \cmark \\
Open-Ended Grading & Manual & \xmark & \xmark & Auto \\
Source-Grounded & \xmark & \xmark & \cmark & \cmark \\
Instant Feedback & MCQ & MCQ & N/A & All \\
Cohort Analytics & Basic & \xmark & \xmark & \cmark \\
Instructor Workflows & \cmark & \xmark & \xmark & \cmark \\
\bottomrule
\end{tabular}
}
\caption{Feature comparison of Edhivo with existing tools. NbLM = NotebookLM.}
\label{tab:comparison}
\end{table}

\section{System Architecture}
\label{sec:architecture}
\begin{figure*}[ht]
    \centering
    \includegraphics[width=\linewidth]{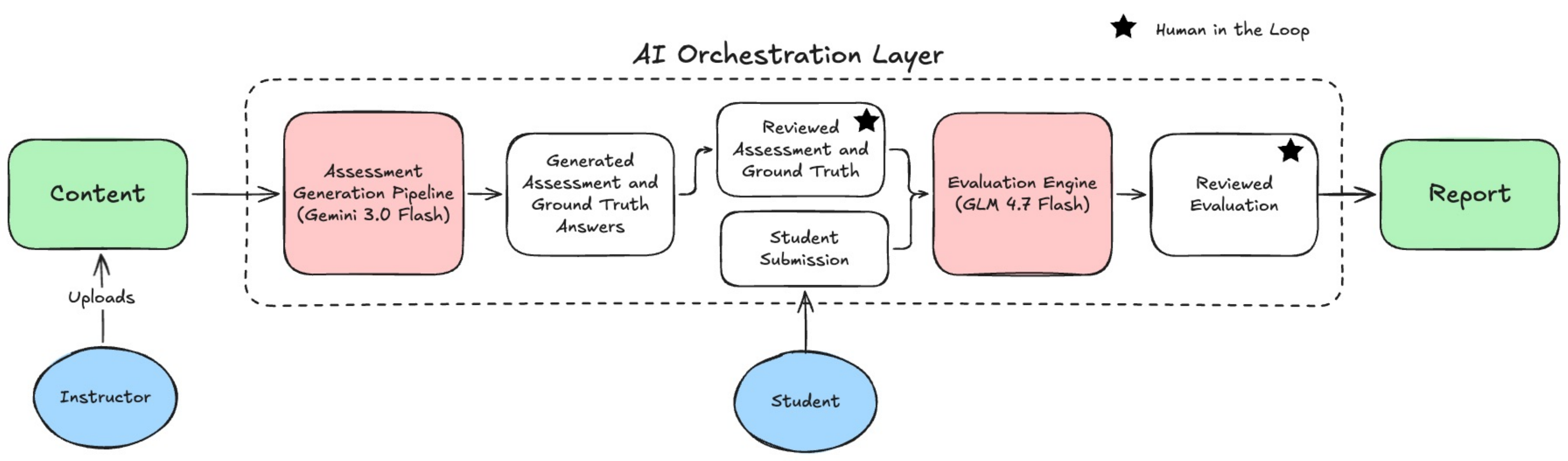} 
    \caption{The Edhivo AI Orchestration Layer Architecture.}
    \label{fig:architecture}
\end{figure*}

The Edhivo architecture is structured as an AI Orchestration Layer (Figure~\ref{fig:architecture}) that oversees the entire lifecycle of an assessment, from content ingestion to student reporting.

\subsection{Model Selection \& Prompt Engineering}
We employ a strategy to use heterogeneous LLM models to optimize for both complex reasoning and system throughput:\footnote{However, the detailed model choices are replaceable.}
\begin{itemize}[leftmargin=*]
    \item \textbf{Assessment Generation:} We utilize Gemini 3 Flash for its 1M+ token context window, allowing the system to ingest high-density course materials (textbooks, transcripts) without truncation.
    \item \textbf{Evaluation Engine:} We deploy GLM 4.7 Flash, which demonstrates high instruction-following capabilities for multi-criteria grading tasks.
\end{itemize}

To maintain a deterministic state between the AI Orchestration Layer and the user interface, we leverage each model's native structured output capabilities. Each request includes a JSON schema that specifies the data types for questions, options, and grading keys.

\subsection{Assessment Generation Pipeline}
\label{sec:generation}
The generation pipeline transforms raw course content into a deployable assessment through a user-configured, parallel generation process. The instructor (or student, in self-study mode) specifies the desired number of MCQs and open-ended questions independently, and the system dispatches parallel requests to the generation model---one per question type---each governed by a dedicated prompt with targeted few-shot examples \cite{brown2020language} to prevent cross-type prompt drift.
Each request follows the prompt template~\ref{lst:gen_template} below, combining a role definition, the ingested course context, user-defined constraints, and high-quality exemplars.

\begin{lstlisting}[frame=single, breaklines=true, basicstyle=\ttfamily\small, columns=fullflexible, caption={Assessment Generation Template}, label={lst:gen_template}]
[Role]: Expert Academic Question Setter.
[Context]: {uploaded_context}
[Constraints]: Generate {n} questions at {level} difficulty.
[Format]: Strictly follow the JSON schema {schema}.
[Examples]: {few_shot_examples}
\end{lstlisting}

By enforcing strict schema adherence, the generated content is immediately ready for Human-in-the-Loop review. Full prompts are provided in Appendix~\ref{sec:appendix}.

\subsection{Rubric-Based Automated Evaluation Engine}
\label{sec:evaluation_engine}
Upon submission, the evaluation model performs a semantic comparison of each student response against the instructor-defined ground truth \cite{zheng2023judging, liu2023geval}. A Chain-of-Thought (CoT) \cite{wei2022chain} grading prompt (Prompt~\ref{lst:grading_prompt}) forces the model to provide step-by-step justifications across three explicit criteria---Correctness, Completeness, and Relevance---before assigning scores. This ``reason-then-score'' design aims to reduce grading arbitrariness and reflects calls for fine-grained, context-aware evaluation \cite{song-etal-2025-prmbench,zhou-2026-position}.

\begin{lstlisting}[frame=single, breaklines=true, basicstyle=\ttfamily\small, columns=fullflexible, caption={Grading \& Justification Prompt (Abridged)}, label={lst:grading_prompt}]
Task: Evaluate the [Student's Answer] against the [Reference Answer].
Evaluation Criteria:
1. Correctness (0-5): Factual alignment.
2. Completeness (0-5): Coverage of key concepts.
3. Relevance (0-5): Directness of response.
Reasoning: Perform a step-by-step analysis of each criterion. 
Output: Return a JSON object with scores and a concise justification.
\end{lstlisting}

\section{Platform Interfaces \& Analytics}
\label{sec:analytics}

\subsection{Instructor Interface: Content-to-Assessment}
\begin{figure}[t]
    \centering
    \includegraphics[width=\linewidth]{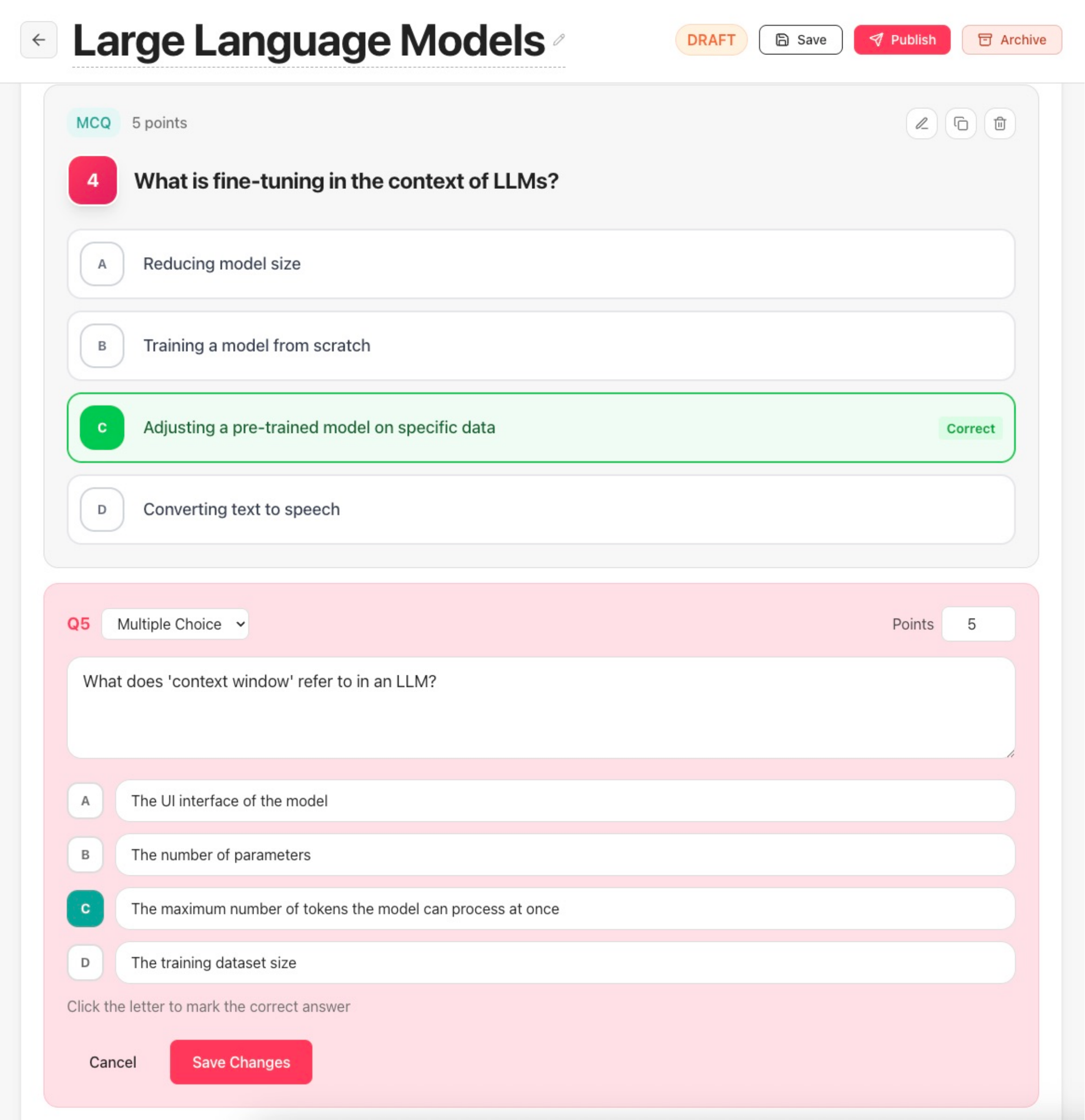} 
    \caption{The instructor review page for the Content-to-Assessment workflow.}
    \label{fig:instructor_review}
\end{figure}

After creating a course and enrolling students, the instructor uploads source materials and the AI Orchestration Layer proposes an assessment. The workflow follows: Upload Content $\rightarrow$ Review Synthetic Questions $\rightarrow$ Revise or Override $\rightarrow$ Publish. Instructors can (1)~edit or replace AI-generated questions, (2)~provide additional steering prompts to guide generation, or (3)~author assessments fully manually without AI assistance. Figure~\ref{fig:instructor_review} shows the instructor review interface.

\subsection{Student Interface: Real-time Feedback Loop}
\begin{figure}[t]
    \centering
    \includegraphics[width=\linewidth]{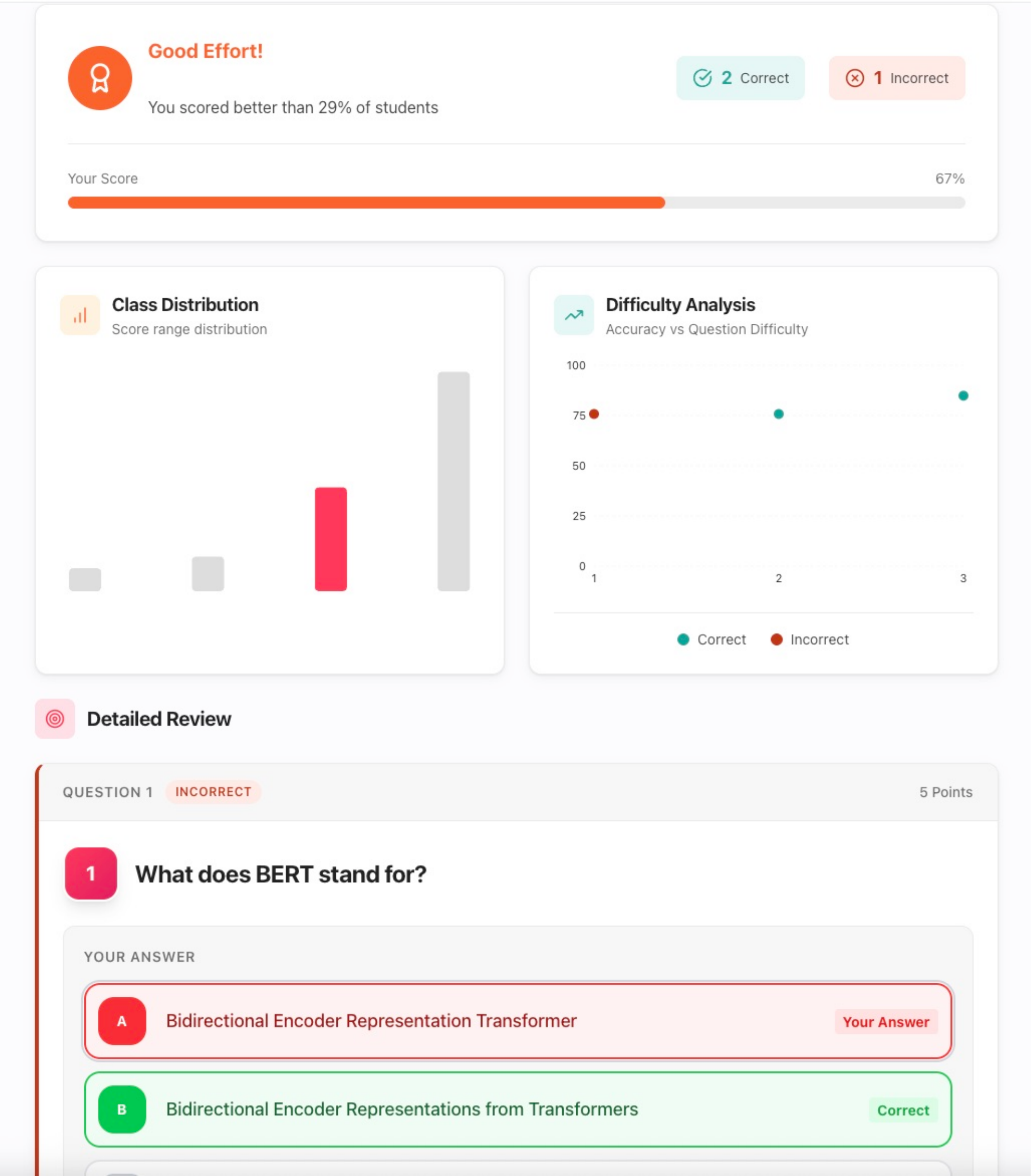} 
    \caption{A student feedback report featuring real-time insights.}
    \label{fig:student_feedback}
\end{figure}

Students interact with a streamlined dashboard providing three key capabilities:
\begin{itemize}[leftmargin=*]
    \item \textbf{Active Learning:} Students can generate self-study assessments and ``propose'' questions to the instructor, crowdsourcing the question bank.
    \item \textbf{Pre/Post Assessment:} The same assessment can be administered before and after learning; the score delta provides a direct measure of learning gain.
    \item \textbf{Instant Remediation:} Upon submission, students receive qualitative feedback from the evaluation engine (Section~\ref{sec:evaluation_engine}), enabling correction of misconceptions while material is fresh \cite{shute2008focus}.
\end{itemize}
Figure~\ref{fig:student_feedback} shows the student feedback report.

\subsection{Hierarchical Analytics \& Reporting}
\begin{figure}[t]
    \centering
\includegraphics[width=\linewidth]{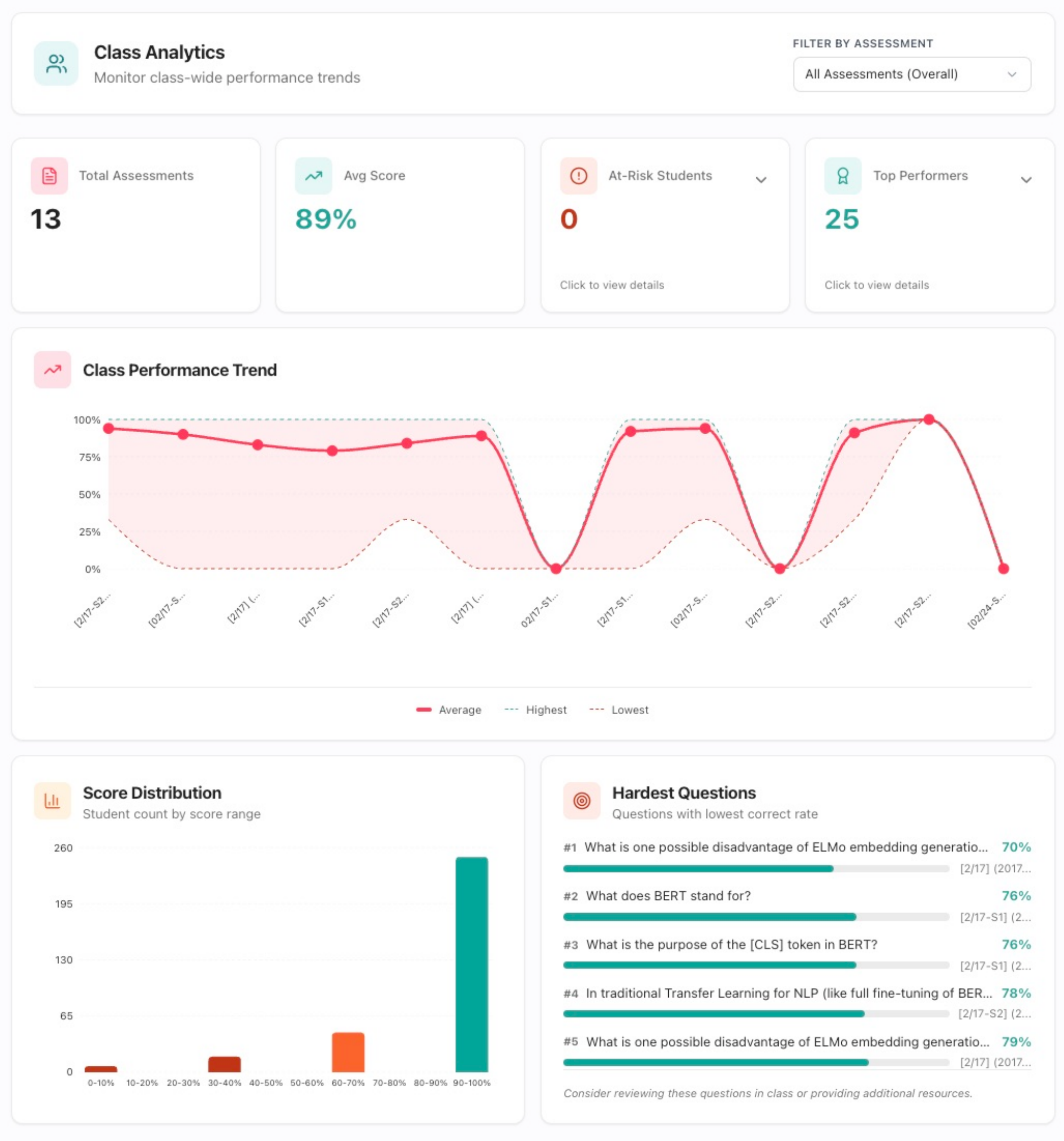} 
    \caption{The class progress report for the instructor.}
    \label{fig:class_progress}
\end{figure}

The reporting engine delivers a ``Pedagogical Health'' snapshot at four levels of granularity: (1)~\textbf{Student-Assessment}---individual scores and feedback per test; (2)~\textbf{Student-Course}---longitudinal trends across modules; (3)~\textbf{Cohort-Assessment}---class-wide score distributions to surface bottleneck concepts; and (4)~\textbf{Cohort-Course}---macro-level progress tracking over the semester. Figure~\ref{fig:class_progress} shows the instructor-facing class progress report.

\section{Pilot Study: Real Classroom Deployment}
\label{sec:pilot-study}

To evaluate its practical feasibility and pedagogical utility, we conducted a pilot deployment in a real graduate-level classroom in Spring 2026. The deployment aimed to assess whether the platform could support high-frequency formative assessment with minimal feedback latency in a live instructional setting.

\paragraph{Course Context}

The platform was deployed in AMS 691.01: Foundations and Frontiers of Large Language Models at Stony Brook University, with 36 graduate students from data science, computer science, and engineering backgrounds. The course covers the evolution of large language models from early Transformer architectures to modern reinforcement-learning-based systems, and includes faculty lectures, student-led research presentations, and project-based components.\footnote{\url{https://joezhouai.com/llm-course-26/}}

\paragraph{Assessment-Centered Pedagogy}

A core instructional design element was the integration of structured pre-/post-assessments into student-led presentations. In prior offerings, student engagement during peer presentations was uneven, and immediate learning outcomes were difficult to observe. In Spring 2026, each presentation incorporated a short assessment (typically three multiple-choice questions) authored by the presenting students. The class completed the assessment immediately before and after the presentation. The score difference between post- and pre-assessments provided a lightweight measure of immediate learning gain.

On average, three such pre-/post-assessment cycles were conducted during each 80-minute session. Although each assessment was brief, the frequency was high and sustained throughout the semester. All assessments were formative and did not contribute to course grades. Implementing this design requires near-instant grading and aggregation, making automation essential.

\paragraph{Deployment Workflow}

Edhivo supported assessment authoring, distribution, grading, and aggregation throughout the semester. Student presenters created assessments within the platform, which were reviewed and published by the instructor. Students completed assessments in class, and results were evaluated automatically upon submission.

Each assessment cycle required approximately three minutes for completion, with immediate availability of individual and cohort-level analytics. The instructor dashboard displayed per-question distributions and pre-/post-comparisons, enabling real-time identification of conceptual bottlenecks and guiding classroom discussion.

Informal qualitative feedback indicated that students perceived the process as lightweight and engaging, and that immediate pre-/post-comparisons reinforced attention during presentations. From the instructor’s perspective, the platform enabled sustained high-frequency formative assessment that would have been logistically impractical without automated evaluation.

\section{Evaluation}
\label{sec:eval}

We gather experience from our real classroom pilot study for both qualitative and quantitative evaluation of Edhivo.

\subsection{System Performance: Quantifying Instructor Time Savings}

\paragraph{Assessment Generation Efficiency}
Manual assessment authoring is a cognitively demanding task. We estimate that an instructor takes an average of 30 minutes to review source material, formulate questions, and draft a corresponding answer key (ground truth).

In contrast, Edhivo's generation pipeline issues parallel requests to the generation model, producing both MCQ and open-ended assessments in approximately 10 seconds (median across pilot runs). Even when we factor in a 5-minute Human-in-the-Loop (HITL) review period for the instructor to refine the AI's output, the net time saved remains a significant 25 minutes per assessment (Table \ref{tab:gen_latency}).

\begin{table}[t]
\centering
\resizebox{\columnwidth}{!}{
\begin{tabular}{lccc}
\toprule
\textbf{Task Phase} & \textbf{Manual (Est.)} & \textbf{Edhivo (AI+HITL)} & \textbf{Time Saved} \\
\midrule
Content Review \& Drafting & 30 min & 10 sec & -- \\
Instructor Review/Refine & -- & 5 min & -- \\
Total Prep Time & 30 min & $\sim$5.2 min & $\sim$25 min \\
\bottomrule
\end{tabular}
}
\caption{Comparison of Assessment Generation Latency.}
\label{tab:gen_latency}
\end{table}

\paragraph{Scalable Evaluation \& Grading}
Manual grading scales linearly with cohort size and is especially burdensome for open-ended questions. Edhivo's MCQ engine grades deterministically and instantly; for open-ended responses, the evaluation model processes the entire cohort asynchronously in approximately 15--20 seconds. Even including a human audit phase (estimated at 2.5 min/student), the platform reduces grading time by over 8 hours for a 200-student assessment (Table~\ref{tab:scaling_lat}).

\begin{table}[t]
\centering
\resizebox{\columnwidth}{!}{
\begin{tabular}{lccc}
\toprule
\textbf{Cohort Size} & \textbf{Manual (Open-Ended)} & \textbf{Edhivo (Grading+Audit)} & \textbf{Net Time Saved} \\
\midrule
30 Students & 150 min (2.5 hrs) & 75 min & 1.25 hrs \\
200 Students & 1,000 min (16.6 hrs) & 500 min & 8.3 hrs \\
\bottomrule
\end{tabular}
}
\caption{Scaling of Open-Ended Evaluation Latency (assuming 5 min/student human baseline).}
\label{tab:scaling_lat}
\end{table}

\paragraph{Cost-Efficiency Analysis}
At OpenRouter pricing for GLM 4.7 Flash, a 10-question assessment costs \$0.00275 per student, meaning a 200-student cohort can be graded for under \$0.55 (Table~\ref{tab:unit_economics}).

\begin{table}[t]
\centering
\resizebox{\columnwidth}{!}{
\begin{tabular}{lcccc}
\toprule
\textbf{Component} & \textbf{Tokens/Q} & \textbf{Price/1M} & \textbf{Cost/Question} & \textbf{Cost/Student (10Q)} \\
\midrule
Input (Q + GT + Rubric + Prompt) & 1,400 & \$0.125 & \$0.000175 & \$0.00175 \\
Output (Justification) & 200 & \$0.500 & \$0.000100 & \$0.00100 \\
\textbf{Total per Student} & 1,600 & -- & \$0.000275 & \$0.00275 \\
Total per 30 Students & -- & -- & -- & \$0.0825 \\
Total per 200 Students & -- & -- & -- & \$0.55 \\
\bottomrule
\end{tabular}
}
\caption{Unit economics of automated evaluation using GLM 4.7 Flash via OpenRouter.}
\label{tab:unit_economics}
\end{table}

\paragraph{Pedagogical Alignment \& Reliability}
Although computational efficacy and cost savings are essential for institutional scalability, grading precision remains our primary architectural concern. To validate the pedagogical integrity of our system, we are assembling a longitudinal dataset that compares AI-determined evaluations with instructor-determined scores \cite{kullman2023survey}. In the interim, the Human-in-the-Loop (HITL) audit process serves a dual purpose:

\begin{enumerate}[leftmargin=*]
    \item \textbf{Safety Mechanism:} It acts as a final fail-safe to prevent model hallucinations from affecting student records.
    \item \textbf{Active Learning Loop:} It functions as a structured data-collection loop. By collecting instructor overrides and modifications, we can fine-tune the Evaluation Engine’s prompt strategy to better align with specific academic standards and departmental rubrics.
\end{enumerate}

\subsection{Student Evaluation and Pedagogical Impact}
To evaluate the perceived efficacy of Edhivo, we conducted a post-pilot survey among the students at Stony Brook University. The survey was sent to 36 students in the pilot study course described in Section~\ref{sec:pilot-study}, and 30 responses were received (83.3\% response rate). The survey focused on two key pedagogical dimensions: Ease of Use and the Impact of Instant Feedback.

\begin{figure}[t]
    \centering
    \includegraphics[width=\linewidth]{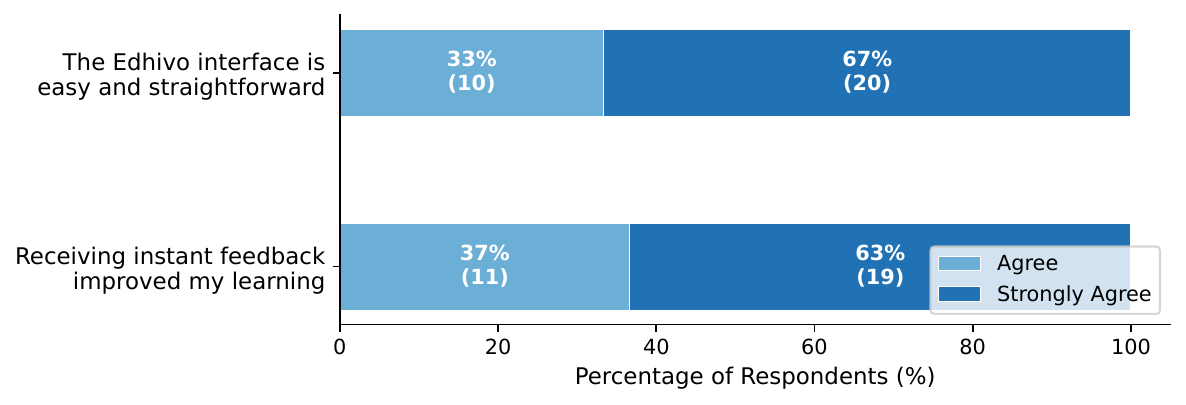}
    \caption{Student agreement on platform usability and learning impact.}
    \label{fig:likert_chart}
\end{figure}

\paragraph{Quantitative Analysis: Likert Distribution}
The quantitative results indicate a strong consensus on the platform's utility (see Figure \ref{fig:likert_chart}).

\begin{itemize}[leftmargin=*]
    \item \textbf{Usability:} 100\% of participants either ``Agreed'' or ``Strongly Agreed'' that the Edhivo interface is easy and straightforward to use.
    \item \textbf{Learning Impact:} 100\% of students ``Agreed'' or ``Strongly Agreed'' that receiving instant feedback improved their learning experience compared to traditional multi-day grading cycles.
\end{itemize}

\begin{figure}[H]
    \centering
    \includegraphics[width=\linewidth]{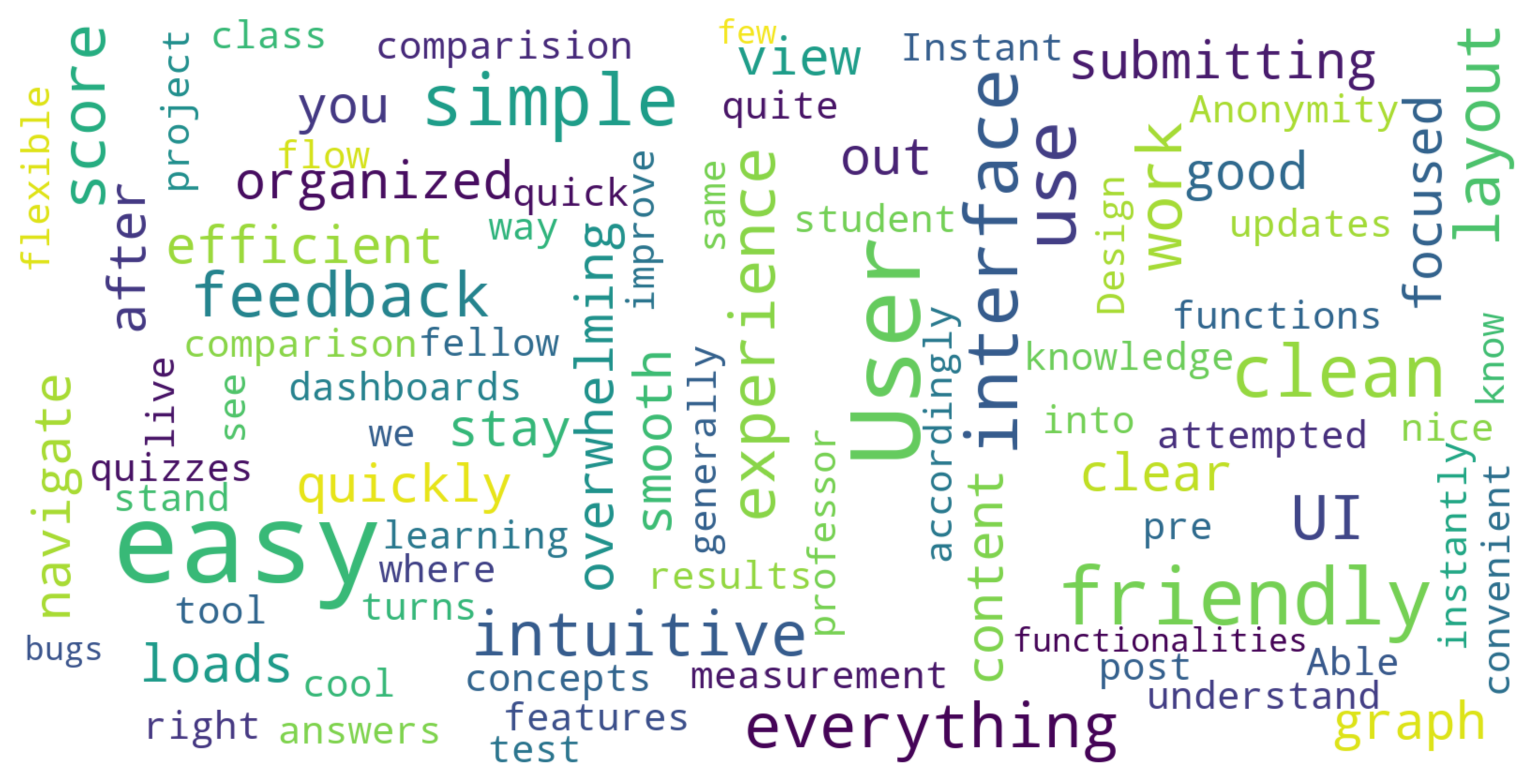}
    \caption{Word cloud representing student sentiment and feature highlights.}
    \label{fig:wordcloud}
\end{figure}

\paragraph{Qualitative Insights: The ``Voice of the Student''}
We collected open-ended feedback to identify specific features that contributed to the student experience. A thematic analysis---represented by the word cloud in Figure \ref{fig:wordcloud}---highlights three dominant themes: User-Friendliness, Intuitiveness, and the Instant Feedback Loop.

These results suggest that by removing the ``administrative friction'' of traditional LMS interfaces and the ``temporal friction'' of manual grading, Edhivo allows students to remain in a state of active cognitive engagement with the course material \cite{hattie2007power}.

\section{Conclusion}
We presented Edhivo, an AI-native pedagogical platform that closes the feedback gap in higher education through a vertically integrated system for content delivery, assessment generation, and automated evaluation. Our pilot deployment at Stony Brook University demonstrates significant reductions in instructor workload and feedback latency at minimal infrastructure cost. Future work will focus on formal Inter-Rater Reliability analysis across larger, multi-disciplinary cohorts and expanding the platform's applicability beyond STEM disciplines.

\section*{Limitations}
While Edhivo demonstrates significant potential in reducing the assessment authoring, management, and evaluation gaps, several limitations remain. 

First, the platform lacks \textbf{proctoring safeguards} such as full-screen enforcement and tab-switch detection, which are critical for high-stakes testing. This is a common challenge in digital assessment, especially as AI advances.

Second, our evaluation engine may be susceptible to linguistic bias; students using LLMs to draft responses might receive higher scores due to stylistic alignment between models \cite{zheng2023judging}, potentially disadvantaging students with non-standard writing styles. 

Third, the domain-specific efficacy of the platform is currently centered on STEM-related subjects with objective ground truths. Further research is required to determine if the rubric-based logic holds for more subjective disciplines such as the humanities. 

Additionally, reported time savings for manual authoring (Section~\ref{sec:eval}) are informal estimates from a single instructor, not controlled measurements across multiple instructors or disciplines; actual authoring times may vary with subject complexity and individual workflows.

Finally, our classroom pilot was limited to a small cohort ($n=36$) at a single institution, and our evaluation data derives from a single course context. Future work will focus on longitudinal studies across larger, multi-disciplinary student bodies to better assess the Inter-Rater Reliability (IRR) between the AI Orchestration Layer and human expert graders.

\FloatBarrier

\section*{Ethics Statement}
This work involves the collection and processing of student assessment data and survey responses in an educational setting. No personally identifiable student data was collected or retained. All survey participation was voluntary, and responses were anonymized prior to analysis.

We acknowledge that AI-based evaluation may introduce linguistic bias, potentially favoring responses stylistically aligned with the evaluation model's training data. This risk is discussed in the Limitations section. The HITL audit process serves as a safeguard to prevent unverified AI scores from impacting official student records.

\newpage

\appendix

\section{Full Prompts}
\label{sec:appendix}

\subsection{Evaluation Prompt: Rubric-Based Grading}
The following prompt is used by the evaluation engine to score student responses against instructor-defined ground truth. The prompt employs Chain-of-Thought reasoning to ensure transparent and consistent grading.

\begin{lstlisting}[frame=single, breaklines=true, basicstyle=\ttfamily\footnotesize, columns=fullflexible]
You are an expert Teaching Assistant grading an open-ended question. Your goal is to provide a fair and consistent evaluation based on a provided rubric.

**Question:**
{question}

**Reference Answer (The ideal answer):**
{ground_truth}

**Student's Answer:**
{given_answer}

---

**Evaluation Task:**
Please evaluate the student's answer against the reference answer using the following rubric. Provide your reasoning step-by-step (chain-of-thought) before giving the final output.

**Rubric:**
1. Correctness (0-5 points): Is the student's answer factually correct and aligned with the reference answer? (5 = Completely correct, 0 = Completely incorrect).
2. Completeness (0-5 points): Does the answer address all key concepts and parts of the question present in the reference answer? (5 = All key concepts covered, 0 = No key concepts covered).
3. Relevance (0-5 points): Is the answer directly addressing the question asked, without irrelevant information? (5 = Perfectly relevant, 0 = Completely irrelevant).

**Chain of Thought (Your reasoning):**
* Correctness Analysis: [Examine factual accuracy...]
* Completeness Analysis: [Identify missing concepts...]
* Relevance Analysis: [Assess directness and focus...]

**Final Output:**
Provide your final scores and overall judgment in a JSON object.
{
  "scores": {
    "correctness": <score_1>,
    "completeness": <score_2>,
    "relevance": <score_3>
  },
  "justification": "<A concise summary of your reasoning>"
}
\end{lstlisting}

\subsection{Generation Prompt: Multiple-Choice Questions}
The following prompt is used to generate context-aware multiple-choice questions from instructor-provided materials.

\begin{lstlisting}[frame=single, breaklines=true, basicstyle=\ttfamily\footnotesize, columns=fullflexible]
You are an expert educational assessment creator specializing in multiple-choice questions. Your task is to generate high-quality MCQ questions that test conceptual understanding rather than rote memorization.

**Source Content:**
{context}

**Generation Requirements:**
- Number of questions: {num_questions}
- Difficulty level: {difficulty_level}
- Focus areas: {focus_areas}

**Quality Guidelines:**
1. Test conceptual understanding and application, not just recall
2. Provide exactly 4 options (A, B, C, D) with one unambiguously correct answer
3. Design plausible distractors based on common misconceptions
4. Ensure each question is self-contained and unambiguous
5. Vary question stems (e.g., "What is...", "Which of the following...", "Why does...")

**Output Format (JSON array):**
[
  {
    "type": "MCQ",
    "text": "Question text here?",
    "options": ["Option A", "Option B", "Option C", "Option D"],
    "correct_answer": "A",
    "explanation": "Brief explanation of why this is correct and why distractors are incorrect"
  }
]

**Few-Shot Example:**
{
  "type": "MCQ",
  "text": "Which data structure provides O(1) average-case time complexity for insertion, deletion, and lookup operations?",
  "options": [
    "Binary Search Tree",
    "Hash Table",
    "Linked List",
    "Array"
  ],
  "correct_answer": "B",
  "explanation": "Hash tables provide constant-time average-case performance for these operations. BSTs provide O(log n), linked lists O(n), and arrays O(n) for insertion/deletion."
}

**IMPORTANT:** Return ONLY the JSON array, no additional text, markdown, or code fences.
\end{lstlisting}

\subsection{Generation Prompt: Open-Ended Questions}
The following prompt is used to generate open-ended questions that require qualitative responses.

\begin{lstlisting}[frame=single, breaklines=true, basicstyle=\ttfamily\footnotesize, columns=fullflexible]
You are an expert educational assessment creator specializing in open-ended questions. Your task is to generate questions that elicit deeper reasoning and require students to articulate their understanding in their own words.

**Source Content:**
{context}

**Generation Requirements:**
- Number of questions: {num_questions}
- Difficulty level: {difficulty_level}
- Focus areas: {focus_areas}

**Quality Guidelines:**
1. Design questions that require explanation, analysis, or synthesis
2. Avoid questions with single-word or yes/no answers
3. Provide a comprehensive reference answer covering all key concepts
4. Include explanation of critical points the answer must address
5. Ensure questions are specific enough to guide response scope

**Output Format (JSON array):**
[
  {
    "type": "TEXT",
    "text": "Open-ended question here?",
    "correct_answer": "Comprehensive reference answer demonstrating full understanding",
    "explanation": "Key concepts and points that must be addressed for full credit"
  }
]

**Few-Shot Example:**
{
  "type": "TEXT",
  "text": "Describe the key differences between supervised and unsupervised machine learning, providing one example application for each.",
  "correct_answer": "Supervised learning uses labeled training data where inputs are paired with correct outputs, allowing the model to learn mappings for prediction tasks (e.g., email spam classification). Unsupervised learning works with unlabeled data to discover hidden patterns or structures without predefined outputs (e.g., customer segmentation through clustering). The key distinction is the presence or absence of labeled training data and whether the goal is prediction versus pattern discovery.",
  "explanation": "Must include: (1) labeled vs unlabeled data distinction, (2) prediction vs pattern discovery goals, (3) valid example for each paradigm, (4) clear contrast between the two approaches"
}

**IMPORTANT:** Return ONLY the JSON array, no additional text, markdown, or code fences.
\end{lstlisting}

\end{document}